\documentclass[aps,prb,amsmath,amssymb,twocolumn,superscriptaddress,showpacs,floatfix]{revtex4-1}
\usepackage{ifpdf}
 \newif\ifpdf
\ifx\pdfoutput\undefined
   \pdffalse
\else
   \pdfoutput=1
   \pdftrue
\fi
\ifpdf
   \usepackage{graphicx}
   \usepackage{epstopdf}
   \usepackage{color}
\else
   \usepackage{graphicx}
\fi
\usepackage{dcolumn}
\usepackage{bm}
\usepackage{srctex}
\usepackage{hyperref}

\DeclareMathOperator{\TCFE}{\mathit{T}_{\mathrm{\scriptscriptstyle C}}^{\mathrm{\scriptscriptstyle FE}}}
\DeclareMathOperator{\DTFE}{\Delta \mathit{T}^{\mathrm{\scriptscriptstyle FE}}}

\DeclareMathOperator{\TCFM}{\mathit T_{\mathrm{\scriptscriptstyle C}}^{\mathrm{\scriptscriptstyle FM}}}

\begin{document}


\title{Competition of magneto-dipole, anisotropy and exchange interactions in composite multiferroics}

\author{A.~M.~Belemuk}
\affiliation{Institute for High Pressure Physics, Russian Academy of Science, Troitsk 142190, Russia}
\affiliation{Department of Theoretical Physics, Moscow Institute of Physics and Technology, 141700 Moscow, Russia}

\author{O.~G.~Udalov}
\affiliation{Department of Physics and Astronomy, California State University Northridge, Northridge, CA 91330, USA}
\affiliation{Institute for Physics of Microstructures, Russian Academy of Science, Nizhny Novgorod, 603950, Russia}

\author{N.~M.~Chtchelkatchev}
\affiliation{Department of Physics and Astronomy, California State University Northridge, Northridge, CA 91330, USA}
\affiliation{L.D. Landau Institute for Theoretical Physics, Russian Academy of Sciences,117940 Moscow, Russia}
\affiliation{Department of Theoretical Physics, Moscow Institute of Physics and Technology, 141700 Moscow, Russia}

\author{I.~S.~Beloborodov}
\affiliation{Department of Physics and Astronomy, California State University Northridge, Northridge, CA 91330, USA}

\date{\today}


\begin{abstract}
We study the competition of magneto-dipole, anisotropy and exchange interactions in composite three dimensional multiferroics.
Using Monte Carlo simulations we show that magneto-dipole interaction does not suppress the
ferromagnetic state caused by the interaction of the ferroelectric matrix and magnetic subsystem.
However, the presence of magneto-dipole interaction influences the order-disorder transition: depending on the strength of
magneto-dipole interaction the transition from the ferromagnetic to the superparamagnetic
state is accompanied either by creation of vortices or domains of opposite magnetization. We show that the temperature hysteresis loop occurs due to non-monotonic behavior
of exchange interaction versus temperature. The origin of this hysteresis is
related to the presence of stable magnetic domains which are robust against
thermal fluctuations.
\end{abstract}

\maketitle

\section{Introduction\label{sec:intro}}

Multiferroic materials are materials with coupled magnetic and electric degrees of freedom.~\cite{Cheong07, Khomskii06,Katsuara05, Mostovoy06,Kimura07} One example of this coupling is due to spin-orbit interaction in certain crystals.
However, this coupling is relatively weak.  Currently there is an
active search for multiferroic materials with strong coupling.~\cite{Ederer05, Kornev07}

One way to strongly coupled magnetic and electric degrees of freedom
is to develop hybrid ferroelectric-ferromagnetic
layered materials where mechanical stress produces strong correlations
between the layers.~\cite{Fiebig05, Scott2006, Spal2007, Sone12}
Recently another promising possibility has been suggested based on granular materials
where small metallic ferromagnetic (FM) grains were embedded into ferroelectric (FE) matrix
or these grains were located in close proximity to the FE substrate.~\cite{Bel2014ME}
The presence of small metallic grains increases
the strength of Coulomb interaction providing the necessary coupling
between the FE and FM degrees of freedom.

An important question is to understand the nature of multiferroic state in granular multiferroic materials.
On the mean-field level the properties of these materials have been understood.~\cite{Bel2014ME}
It was shown that the exchange coupling $J$ depends on the properties of the FE matrix/substrate,
in particular on the dielectric permittivity $\epsilon$ of the
surrounding medium.~\cite{Bel2014ME,Bel2014ME1} Due to temperature dependence
of dielectric permittivity, $\epsilon(T)$,  the exchange interaction depends non-monotonically on temperature leading
to the inverse phase transition with paramagnetic phase appearing at lower temperatures
compared to the ferromagnetic phase. The qualitative
behavior of $J$ is different for small and large (compared to the grain size $r_{\rm gr}$)
inter-grain distances $a$: for large inter-grain distances, $a > r_{\rm gr}$, the $J$-value is increased in the vicinity of the FE Curie point due to suppression of the Coulomb blockade effects leading to a different magnetic state at these temperatures.

However, it is still an open question whether the
magneto-electric coupling obtained in the mean-field theory is robust against the
magneto-dipole and anisotropy interactions neglected
in the mean-field approach. We use the numerical modelling to address this question.

We investigate the non-equilibrium meta-stable states in granular system and study
the nature of new meta-stable phases appearing in the system with temperature dependent exchange interaction.
In addition, we answer the
question if temperature dependent exchange interaction can lead to unusual blocking effects.

To be more specific, we study the magnetic behavior of composite multiferroics
with $\TCFE<\TCFM$, where $\TCFE$ is the FE Curie temperature of paraelectric-ferroelectric transition of
the FE matrix and $\TCFM$ is the Curie temperature of FM grains.
We focus on  the temperature range $T \ll \TCFM$ where all grains are in the FM state and study
the phase diagram of granular multiferroics beyond the mean-field approximation
using Monte-Carlo simulations. In particular, we study the combined effect of magnetic anisotropy,
the long-range magneto-dipole (MD) interaction and the exchange coupling.
We show that there is an inverse magnetic transition in the system which is
robust with respect to the MD interaction. This inverse transition disappears
only for strong MD interaction, stronger than the exchange coupling.
In addition, we show that non-monotonic temperature behavior of
intergrain exchange interaction leads to a new type of hysteresis
in composite multiferroics.

Three-dimensional (3D) nanostructures composed of single-domain ferromagnetic
particles has been intensively studied both experimentally and theoretically.~\cite{Twardowski2013,Bertram2001,Chen1991,Parkin1989,Chien1988,Bedanta09,Maylin2000,Trohidou1998,Kleemann2004}
The interplay of magnetic anisotropy, long-range magneto-dipole interaction and short range
exchange interaction defines the magnetic state of the system.
Depending on the ratio of these interactions different magnetic states are possible
in granular ferromagnets.~\cite{Freitas2001,Freitas2006} Among them are superparamagnetic (SPM),
super spin-glass (SSG) and superferromagnetic (SFM) states.

The most studied situation is related to the case of large intergrain distances ($ \geq 2$nm)
and small exchange interaction where magnetic state is defined by the competition
of MD interaction and anisotropy.~\cite{Berkowitz2000, Trohidou1998, Grady1998}
The magnetic anisotropy is responsible for ``blocking'' phenomena and defines the
blocking temperature $T_{\mathrm b}$.~\cite{Chien1988,Chien1991} The weak MD
interaction modifies the blocking temperature,
while the strong MD
interaction leads to the SSG state.~\cite{Ayton95, Ravichandran96, Mamiya99, Djurberg97, Sahoo03}

For small distances between grains ($\sim 1$ nm) the exchange interaction is crucial. It
leads to the formation of the SFM state.~\cite{Sobolev2012,Freitas2001,Hembree1996,Lutz1998,Freitas2007}
In such systems the SPM-SFM transition occurs. Even a weak exchange interaction can
influence the magnetic state of the system.~\cite{Hembree1996} In particular, the FM ordering with long range ferromagnetic correlations appears.

Different theoretical methods have been used to study granular magnets. The mean-field approach
allows to study granular systems with finite short range exchange interaction and zero (or weak)
MD interaction
neglecting fluctuation effects.~\cite{Beattie1990,Munakata2009} Modelling based on Landau-Lifshitz
equation allows to consider the MD, magnetic anisotropy and exchange interactions.~\cite{Ziemann2004}
However, this approach needs to be generalized in the presence of thermal fluctuations by
introducing the Langevin forces.~\cite{Lazaro1998}
These fluctuations are important for granular magnetic systems
since the granular magnetic moment is relatively small and fluctuation effects are pronounced especially near the phase transitions.
However, the inclusion of Langevin forces in the nonlinear spin dynamics is not
numerically efficient. Therefore we use Monte-Carlo (MC) simulations which allow
to study phase transitions in composite multiferroics with strong long-range MD interaction and arbitrary thermal fluctuations.~\cite{Levanyuk1993,Chantrell1994,Trohidou1998, Grady1998,Freitas2006,Bertram2001, Asselin2010PhysRevB,Chantrell2010PhysRevB,Cheng2010, Twardowski2013}

MC modelling strongly depends on the degree of  anisotropy. At strong anisotropy the
problem reduces to the Ising model with magnetic moment of each particle having only two directions
defined by the anisotropy axis. In this case the MC modelling is very efficient and is
based on the trial spin flips. Another type of MC modelling is based on the Heisenberg
model with arbitrary magnetization direction. This model is more general but it requires more
computational time due to spin rotations over the sphere rather than spin flips.~\cite{Matsubara1993,Hinzke1999,Asselin2010PhysRevB,Chantrell2010PhysRevB} We use our own
MC code with random spin-flips and random spin-rotations that is valid for any anisotropy.

The paper is organized as follows:
In Sec.~\ref{results} we formulate our main results. In Sec.~\ref{Sec:Model} we discuss the
model of composite multiferroics. In Sec.~\ref{quantities} we introduce important
physical quantities which we calculate. We discuss our results in
Sec.~\ref{Sec:Results}. The details of our numerical calculations are presented
in Appendix~\ref{App:CalcProc}.

\section{Main results}
\label{results}

Here we summarize our main results. The non-monotonic temperature dependence of exchange interaction
in composite multiferroics leads to the unusual evolution of the magnetic state with temperature.
The intergrain exchange interaction has either peak or deep in the vicinity of the FE phase
transition due to coupling of electric and magnetic degrees of freedom.~\cite{}
In the mean field approximation the peak in the exchange interaction leads to the
onset of FM state in the vicinity of FE phase transition. The deep in the exchange interaction
suppresses the FM state in the vicinity of the FE Curie point.
We use Monte-Carlo simulations to investigate the influence of long-range MD
interaction and magnetic anisotropy on the magnetic phase diagram of composite multiferroics.
We show that MD interaction and anisotropy do not suppress the magneto-electric coupling
in these materials, however their interplay produces a new type of hysteresis.
Our results are the following:

1) The Monte-Carlo simulations reproduce the mean field results in the absence of
MD interaction and magnetic anisotropy. Similar to the mean field approach,
the FM state exists in the vicinity of the FE Curie point and the disordered
state appears away from this region.

2) The finite MD interaction does not suppress the FM ordering in the
vicinity of the FE phase transition even if the MD
interaction is twice stronger than the exchange interaction. The presence
of MD interaction leads to the appearance of the domain structure and
to splitting of uniform FM state. This result is similar to Ref.~\onlinecite{Hembree1996}, where
a weak FM interaction leads to the formation of
FM domains, while strong MD interaction produces a vortex structure.

3) The magnetic state depends on the strength of MD interaction
outside the FM region: the system is in the SPM state
for weak MD interaction and in the antiferromagnetic stripe phase
for strong MD interaction.

4) The magnetic anisotropy does not influence the FM state.
However, it prevents the formation of vortices in the transition region
and leads to a widening of the FM domain.

5) The ``blocking phenomenon'' does not appear at finite magnetic anisotropy and zero MD interaction at considered temperatures meaning that the system has enough time to reach the ground state such that the non-equilibrium state does not appear.

6) The    ``blocking phenomenon'' appears at finite magnetic anisotropy and finite
MD interaction. The temperature hysteresis loop occurs due to non-monotonic behavior
of exchange interaction vs. temperature. The origin of this hysteresis is
related to the presence of stable magnetic domains which are robust against
thermal fluctuations, MD, exchange, and anisotropy interactions.

7) The AFM stripes appear in the case of deep in the exchange interaction.

Below we discuss these results in details.

\section{The model}
\label{Sec:Model}

\subsection{Magnetic subsystem}

In this subsection we consider the magnetic subsystem.
We model a composite multiferroic as an ensemble of FM grains embedded into FE matrix.
All grains are homogeneously magnetized single domain FM particles of the same volume $V$ and
saturation magnetization $M_s$. For temperatures $T \ll \TCFM$, the saturation magnetization $M_s$
is constant. Each grain with volume $V$ has a magnetic moment $\mu= M_s V$ and is treated as a point dipole
located at the centre of the grain. The grains are pinned to the sites of the
regular cubic lattice with lattice spacing $a$ and can freely rotate their adjusting magnetic moments.

The whole system is modelled as a $3D$ lattice of classical spins,
with magnetic moment of the $i$th grain being ${\bm \mu}_i= \mu {\bf S}_i$, where
the unit vector ${\bf S}_i= (S_i^x, S_i^y, S_i^z)$ is the spin of $i$th particle representing
the direction of the magnetic moment.

We assume that each grain has a uniaxial anisotropy.
Spatial distributions of anisotropy axes varies in different experiments and depends on the preparation condition.
The anisotropy axes can be homogeneously distributed over the solid angle, or uniformly distributed in a certain plane.
We assume that the easy axes of all grains are oriented in the z-direction. This
situation is realized in experiment with magnetic field applied during the sample preparation.~\cite{Freitas2007}

The Hamiltonian of the system has the form
\begin{equation} \label{ham}
{\cal H}= {\cal H}_{\rm exc}+ {\cal H}_{\rm dip}+ {\cal H}_{\rm an}.
\end{equation}
The first term, ${\cal H}_{\rm exc}$, describes the exchange coupling between grains $i$ and $j$
\begin{equation} \label{exc}
{\cal H}_{\rm exc}= - J \sum \limits_{\langle i, j \rangle} {\bf S}_i \cdot {\bf S}_j,
\end{equation}
where the sum is over the nearest neighbour pairs of grains.

The second term, ${\cal H}_{\rm dip}$, in Eq.~\eqref{ham} describes the long-range magneto-dipole (MD)
interaction between magnetic moments ${\bm \mu}_i$ and ${\bm \mu}_j$ of individual grains
\begin{equation} \label{dip}
{\cal H}_{\rm dip}=g \sum \limits_{i < j} {\,} \frac{{\bf S}_i \cdot {\bf S}_j {\,} r^2_{ij}- 3({\bf S}_i \cdot {\bf r}_{ij}) ({\bf S}_j \cdot {\bf r}_{ij})}{r^5_{ij}},
\end{equation}
where ${\bf r}_{ij}$ is the distance between magnetic moments at sites $i$ and $j$ measured
in units of lattice spacing $a$ and $g$ is the MD interaction constant.

The third term, ${\cal H}_{\rm an}$, in Eq.~\eqref{ham} describes uniaxial anisotropy energy
\begin{equation} \label{haman}
{\cal H}_{\rm an}= -K \sum_i ({\bf e}_z \cdot {\bf S}_i)^2,
\end{equation}
where $K$ is the temperature independent magnetic anisotropy energy of a single grain.
The unit vector ${\bf e}_z$
defines the direction of the anisotropy easy axis.

We consider the energy parameters $(J, g, K, T)$ in arbitrary units.
Parameters, $g$ and $K$ depend on a grain volume and
can be controlled by varying the grain size.
The dipole coupling $g$ is additionally depends on
the lattice spacing $a$. This allows one to vary parameters $g$
and $K$ in a wide range. The exchange interaction is proportional
to the grain surface and scales with the volume as $V^{2/3}$.
Moreover, the ratio of MD and exchange interactions can be controlled by varying
the interparticle distance $a$. The MD interaction decays as $1/a^3$ with distance, while
exchange interaction decays exponentially $e^{-\kappa a}$, where $\kappa$
is the inverse length which depend on the
band structure of the surrounding FE matrix and the Fermi energy of electrons inside grains.

\subsection{Ferroelectric subsystem}

The FE matrix is characterized by the Curie temperature $\TCFE$.
We study the temperature region in the vicinity of $\TCFE$. The
most important characteristic of the FE matrix in our consideration is the
FE dielectric permittivity $\epsilon$ which has a peculiarity in the vicinity of the phase
transition point, $\TCFE$.~\cite{Levan1983} This peculiar behavior of $\epsilon$
provides a strong coupling between electric and magnetic degrees of freedom.~\cite{Bel2014ME,Bel2014ME1}

\subsection{Interaction of magnetic and ferroelectric subsystems}

Recently the intergrain exchange interaction, $J$, in composite
multiferroics  was studied  and its temperature behavior was predicted.~\cite{Bel2014ME,Bel2014ME1}
In the vicinity of the FE phase transition the exchange interaction has a peculiarity. Depending on the
system parameters it has either peak or deep.
Such peculiarity appears due to combine influence of Coulomb blockade effects and
the temperature dependence of dielectric permittivity of the FE matrix on the exchange interaction.
The peculiarity of the exchange interaction in the vicinity of the
FE Curie point $T=\TCFE$ is related to the peculiarity in the dielectric permittivity $\epsilon$
of the FE matrix. The exchange interaction as a function of $\epsilon$ has the form~\cite{Bel2014ME}
\begin{equation}\label{Eq_J_peak}
J(T)=J_{0}\epsilon^{\gamma a/r_{\rm gr}- 1},
\end{equation}
where parameter $J_0>0$ is the $\epsilon$-independent part of exchange interaction,
$J_0$ decays exponentially with intergrain distance $a$;
$\gamma$ is the numerical coefficient of order one. The dielectric permittivity $\epsilon$
has a peak at temperature $T=\TCFE$. For $\gamma < r_{\rm gr}/a$ (small intergrain distances)
the exchange interaction has a deep,
while for $\gamma > r_{\rm gr}/a$ (large intergrain distances) it has a peak.

The dependence of the intergrain exchange interaction $J$ on the FE
permittivity $\epsilon$ is the signature of magneto-electric coupling emerging in
composite multiferroics. The peak of exchange interaction leads to the unusual magnetic phase diagram:
the FM state appears in the vicinity of the FE Curie temperature $\TCFE$, while
away from $\TCFE$ the system is in the SPM state.
Thus, the FM state in the system exists in a finite region around FE Curie point, $\TCFE$.
In case of deep in the temperature dependence of exchange constant $J$
the opposite situation occurs: the FM state is suppressed in the vicinity of $\TCFE$
due to interaction of magnetic and FE subsystems.

Above effects have been studied in Ref.~\onlinecite{Bel2014ME} using the
mean field approximation without taking into account the MD interaction and anisotropy.
Here we take into account both interactions and study the influence of MD
interaction and magnetic anisotropy on the magneto-electric coupling in composite multiferroics.

For large inter-grain distances, $\gamma > r_{\rm gr}/a$,
the exchange interaction $J(T)$ has the peak. In this work we model this peak as follows
\begin{equation} \label{Jmfrr}
J(T)= J_0 e^{-(T- \TCFE)^2/w^2},
\end{equation}
where $w$ is the width of the exchange peak and $J_0$ is the amplitude
of intergrain exchange interaction. The peak in $J(T)$
occurs at temperatures $T = \TCFE$, when the Coulomb blockade
is suppressed and the electron wave functions become weakly localized leading to a
strong overlap and strong exchange interaction.~\cite{Bel2014ME} The Coulomb blockade
is suppressed when permittivity $\epsilon(T)$ reaches its maximum value at $T=\TCFE$.

For small inter-grain distances, $\gamma< r_{\rm gr}/a$, the
exchange interaction $J(T)$ has a deep which we model as follows
\begin{equation}\label{Eq_J_pit}
J(T)= J_0 \left( 1- e^{-(T- \TCFE)^2/w^2} \right).
\end{equation}

\section{Calculated quantities}
\label{quantities}

In this section we introduce physical quantities which we calculate.
The calculation procedure is described in Appendix~\ref{App:CalcProc}.
The first quantity we calculate is the average magnetization~\cite{Lau89, Holm93, Peczak91, Chen93}
\begin{equation} \label{magn_part}
M(T)=\left| \frac1N \sum \limits_{i= 1}^N {\bf S}_i \right|,
\end{equation}
where $N$ is the number of lattice sites.

In the presence of MD interaction and zero
external magnetic field, the average magnetization $M(T)$ is not an efficient quantity:
the lattice spins form complex magnetic patterns, either domains or vortices, and the mean
magnetization vanishes,
$1/N \sum_{i} {\bf S}_i= 0$, even if locally the system is in the FM state.
To account for local FM correlations we introduce the cell averaged
magnetization, $m(T)$, over a cell with linear size $L_c$. We find that the optimal
size for such cell average is $L_c=5$
\begin{equation} \label{mcell}
m(T)= \left \langle\left|\frac{1}{N_c} \sum \limits_{i} {\bf S}_{i}\right| \right \rangle,
\end{equation}
where summation is over the all grains in a given cell $N_c=L_c^3$ and
the averaging is defined as summation over the all possible positions of the cell
centre, $\langle\rangle=N^{-1}\sum_{i}$.

Next, we introduce  the spin-spin correlation function $G$ as an averaged
correlation function for nearest-neighbour pairs
\begin{equation} \label{Cg}
G= \frac1N \sum \limits_{\bf R} \frac16 \sum \limits_{\bf g} \langle {\bf S}_{\bf R} \cdot {\bf S}_{\bf R+g} \rangle=
\frac{1}{3N} \sum \limits_{\langle i,j \rangle} \langle {\bf S}_i \cdot {\bf S}_j \rangle,
\end{equation}
where ${\bf g}$ are the six vectors of nearest neighbours and the last
summation ${\langle i,j \rangle}$ is over the all nearest-neighbour pairs in the lattice.
The correlation function $G$ is important for understanding magneto-transport in granular magnets.
The magneto-resistance (MR) of granular magnetic film is proportional to this
correlation function, MR(T)$\sim G$. The MR measurements
can be considered as the probes of the magnetic state of the system.

\section{Discussion of results}\label{Sec:Results}

\subsection{Influence of magneto-dipole (MD) interaction on the magnetic phase diagram of composite multiferroics}

In this subsection we discuss the influence of MD interaction on the magnetic phase
diagram of composite multiferroics for the case of large inter-grain distances,
where exchange interaction $J$ has a peak around $\TCFE$, see Eq.~(\ref{Jmfrr}).
\begin{figure}
\includegraphics[width=1\columnwidth]{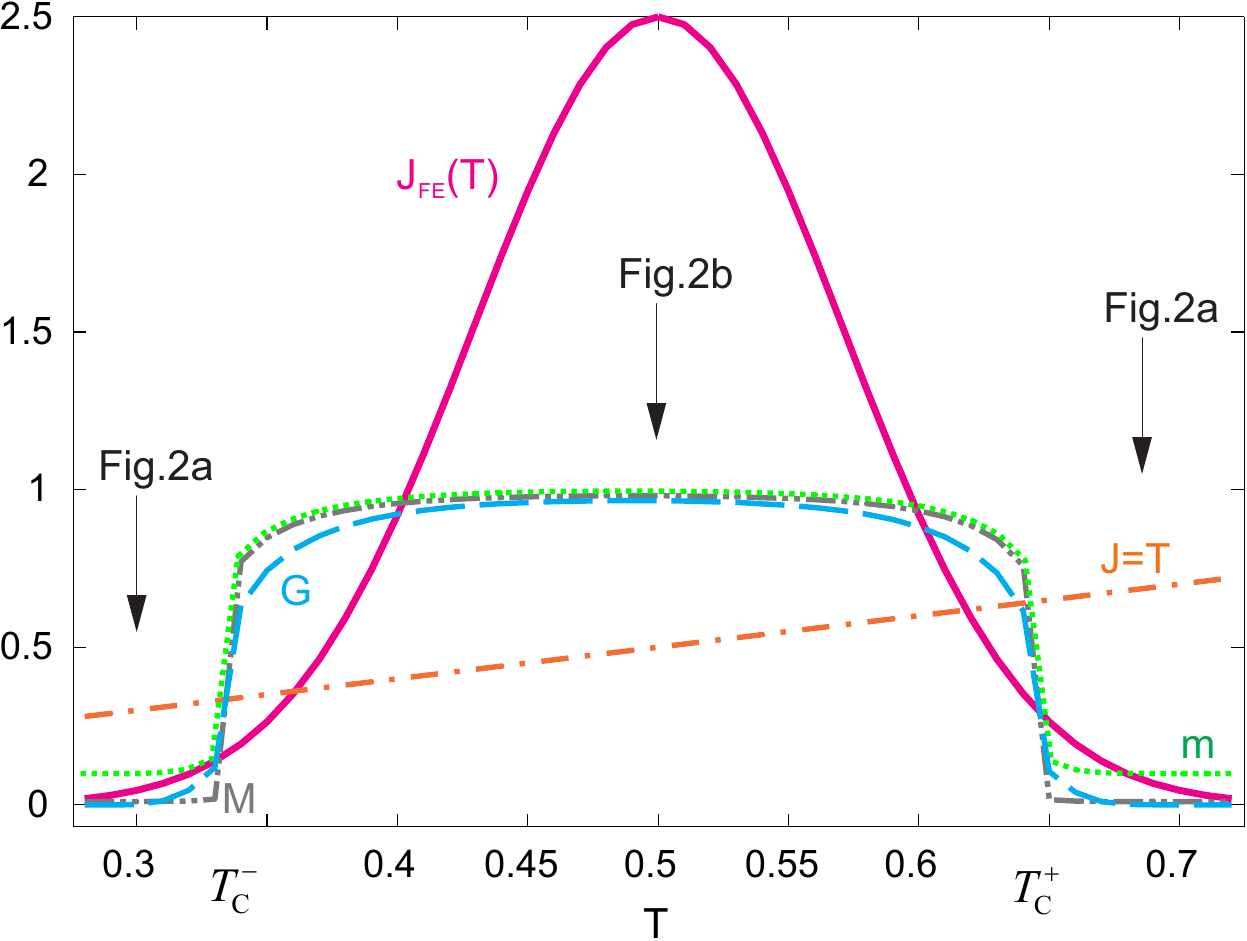}
\caption{(Color online) Magnetic phase diagram
of composite multiferroic vs. temperature for zero magneto-dipole interaction ($g=0$) and magnetic anisotropy ($K=0$).
Solid (red) line shows the temperature dependence of intergrain exchange interaction, $J(T)$. Straight dash dotted (orange) line stands for temperature $T$. Gray dash dot-dotted and green dotted lines show the average magnetization $M(T)$ and cell averaged magnetization $m(T)$, respectively. Blue dashed line shows the nearest neighbour correlation function $G$. Transition temperatures
$T_{\mathrm{\scriptscriptstyle C}}^{\pm}$ are defined using the mean-field equation, $T=2J(T)$.}\label{Fig_PU_0_0}
\end{figure}
\begin{figure}
\includegraphics[width=1\columnwidth]{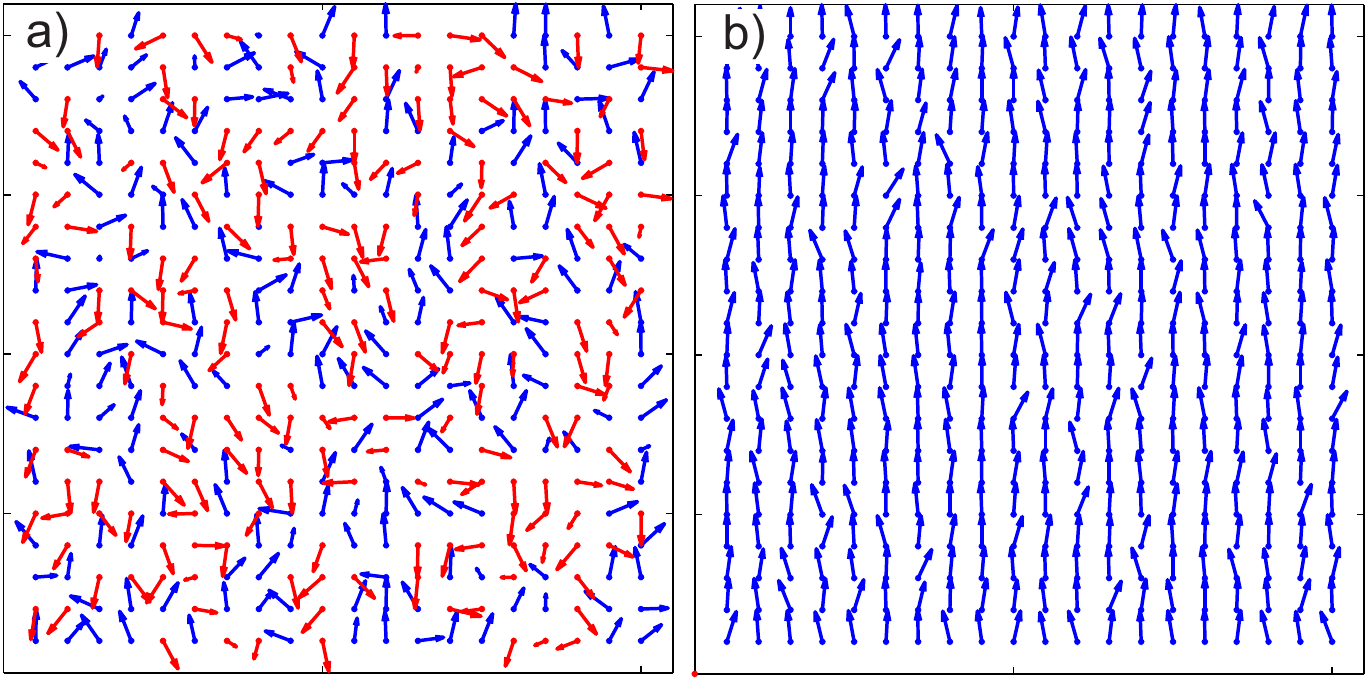}
\caption{(Color online) Snapshots of single magnetic layer of composite multiferroic.
Panels (a) and (b) show the disordered (SPM) magnetic state and the
ordered FM state, respectively. Position of these panels is shown in Fig.~1 by black arrows.}\label{Fig_DisOr}
\end{figure}

\subsubsection{Zero magneto-dipole (MD) interaction}

In the absence of long-range MD interaction and anisotropy the magnetic phase diagram obtained using the Monte-Carlo simulations coincides with the mean-field phase diagram (see Fig.~\ref{Fig_PU_0_0}). We used the following parameters: $J_0=2.5$, $\DTFE=0.1$, $\TCFE=0.5$.
These parameters correspond to Fe grains embedded into organic TTF-CA FE matrix.
The grain size is $5-10$ nm and the intergrain distance  $ \geq 1$ nm. For these parameters the
intergrain exchange interaction is about $J \sim 300$K.
The Curie temperature of bulk ferroelectric TTF-CA is $80$ K. However,
for granular array it can be smaller; the Curie temperature of TTF-CA in
composite granular metal/FE system is $50$ K.~\cite{Keller2014}

The FM state appears around $\TCFE$, where $J(T)$ has a maximum
(see snapshot in Fig.~\ref{Fig_DisOr}b) and the SPM state exists outside this region (see Fig.~\ref{Fig_DisOr}a).
The finite FM region appears with two magnetic phase transitions due to peak in the exchange interaction. In the mean-field approximation
both transition temperatures $T_{\mathrm{\scriptscriptstyle C}}^{\pm}$ are defined as $T=2J(T)$.
Due to zero MD interaction domains are not formed, the ground state of the system is
the homogeneous FM state and the magnetization $M(T)$ is coincide with the cell averaged magnetization $m(T)$.
In the SPM state the saturation magnetization and the correlator $G$ tends to zero since the system is in the disordered state due to thermal fluctuations.
\begin{figure}
\includegraphics[width=1\columnwidth]{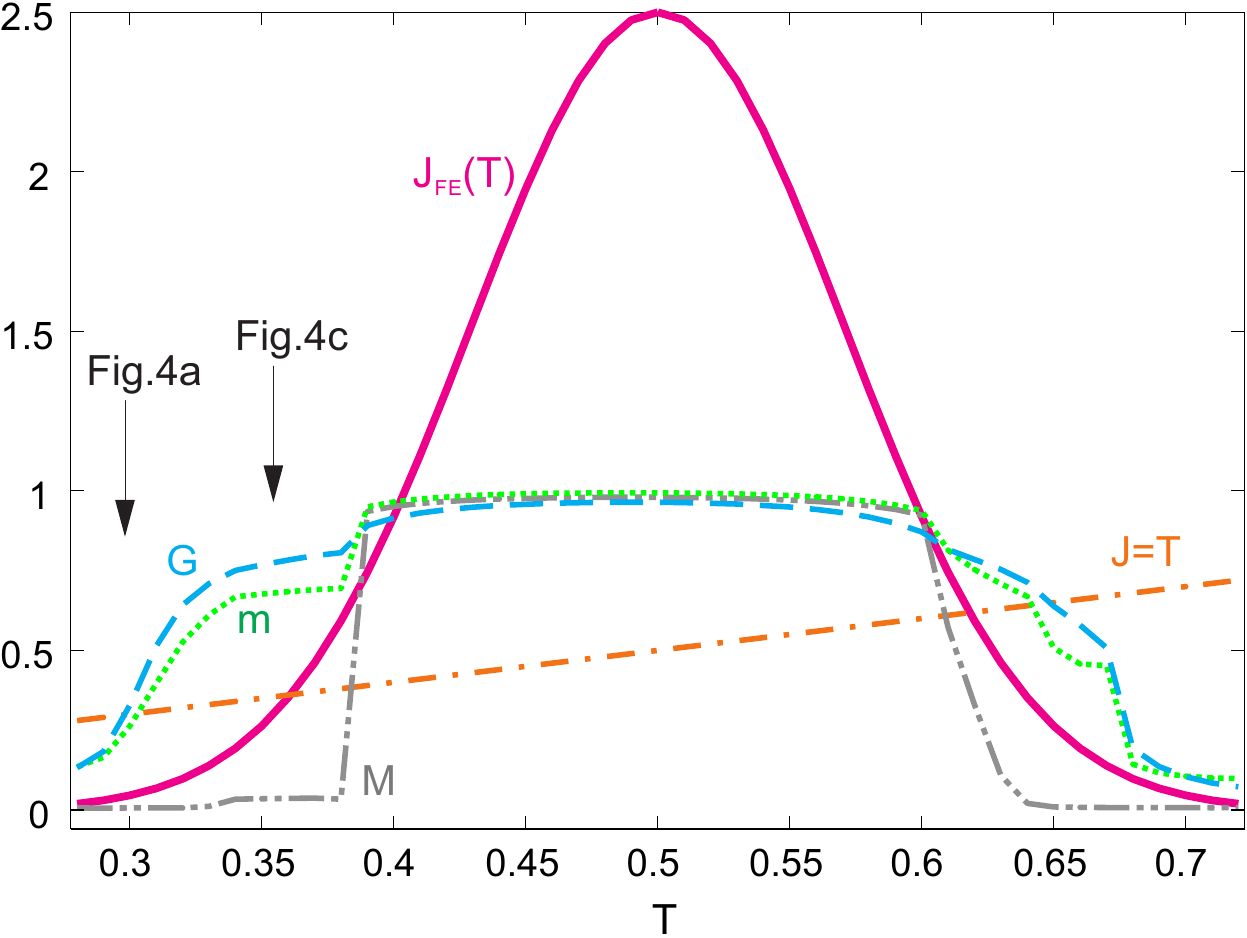}\\
\caption{(Color online) Magnetic phase diagram of composite multiferroic vs. temperature $T$ for weak
magneto-dipole interaction ($g=0.5$) and zero magnetic anisotropy ($K=0$).
All notations are defined in Fig.~1.}\label{Fig_PU_05_0}
\end{figure}
\begin{figure}
\includegraphics[width=1\columnwidth]{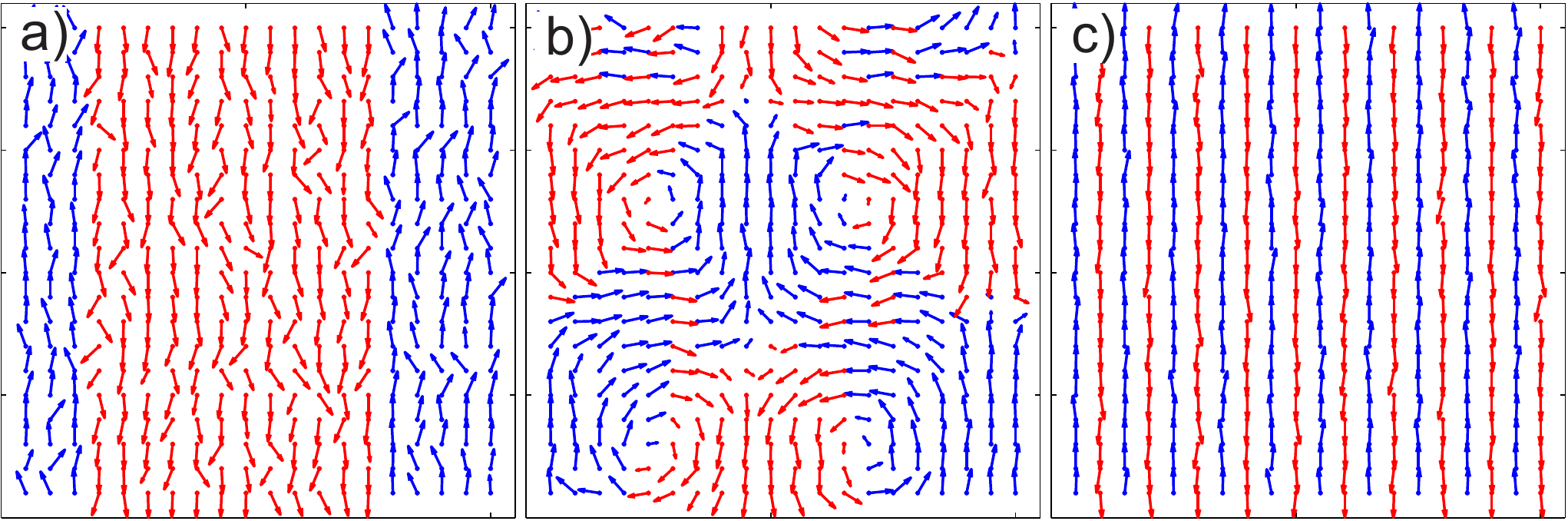}
\caption{(Color online) Snapshots of single magnetic layer of composite multiferroic.
Panel (a) shows the FM state divided into stripe domains appearing
due to the interplay of exchange and weak MD interactions. Panel (b) shows
the vortex state with magnetic vortices appearing due to strong MD interaction.
Panel (c) shows FM chains forming antiferromagnetic pattern appearing
at low temperatures and MD interaction being stronger than exchange interaction.
Position of panels (a) and (c) is shown in Fig.~3 by black arrows.}\label{Fig_DomVor}
\end{figure}

\subsubsection{Weak and moderate magneto-dipole (MD) interaction}

The long-range MD interaction competes with intergrain exchange interaction
suppressing the FM state in the system. Figure~\ref{Fig_PU_05_0} shows the case of weak
MD interaction with the dipole constant, $g=0.5$.
This value of $g$ is typical for Fe grains with size $a=4$ nm, where $g=(2.2 \mu_{\mathrm B}V/\lambda^3_{\mathrm{Fe}})^2/a^3\approx50$ K, $\lambda_{\mathrm{Fe}}=0.28$ nm is the Fe lattice parameter.~\cite{Chelikowsky2006} For these parameters
the FM region exists.
However, the MD interaction reduces the size of FM state
and leads to the formation of domains in the system (see the left panel in Fig.~\ref{Fig_DomVor}).
Above the FM region the SPM state appears, similar to the case of zero MD interaction,
meaning that the thermal fluctuations exceed the exchange and MD interactions.
Below the FM region the thermal fluctuations exceed the exchange interaction but not
the MD interaction. As a result the antiferromagnetic stripes appear at low temperatures, $T < 0.35$
(see panel (c) in Fig.~\ref{Fig_DomVor}). The transition to stripe structure occurs
via formation of large antiferromagnetic domains with temperature dependent sizes.
\begin{figure}
\includegraphics[width=1\columnwidth]{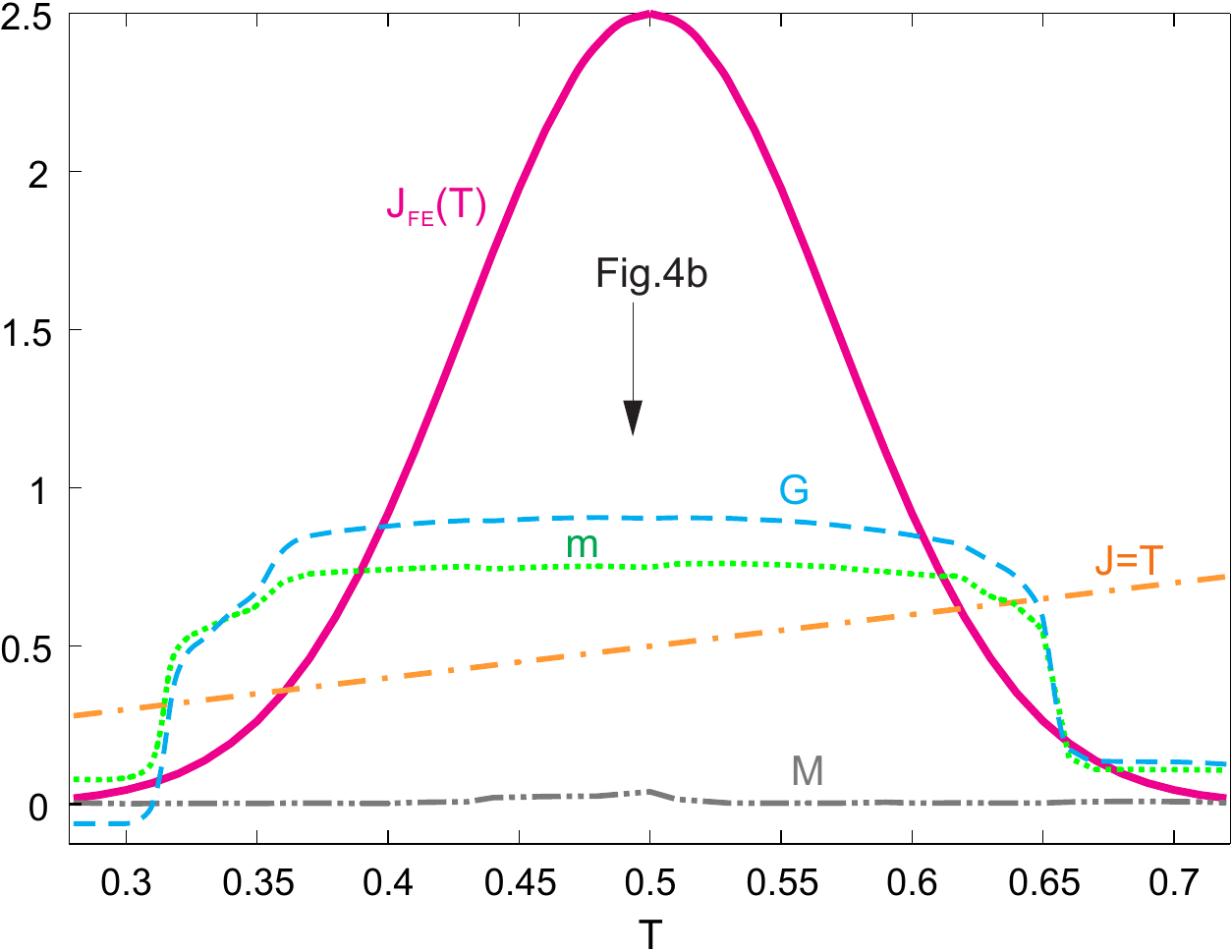}\\
\caption{(Color online) Magnetic phase diagram of composite multiferroic vs.
temperature $T$ for intermediate magneto-dipole interaction $g=2.0$ and zero magnetic
anisotropy ($K=0$). All notations are defined in Fig.~1.}\label{Fig_PU_2_0}
\end{figure}

\subsubsection{Strong magneto-dipole (MD) interaction}

For strong MD interaction, $g\geq 2.0$, the uniform FM state does not appear
in the system (see Fig.~\ref{Fig_PU_2_0}).
However, the long-range magnetic correlations are still exist
in the system close to $\TCFE$.
For such a strong MD interaction the vortex-like structure appears
in the system (see the central snapshots in Fig.~\ref{Fig_DomVor}(b)).
The vortex structure transforms into the stripe structure outside the $\TCFE$ region.
For strong MD interaction the SPM state does not
appear for temperatures above and below $\TCFE$,
instead the stripe structure appears in these regions.
At higher temperatures the stripe structure transforms into SPM state
due to thermal fluctuations.

The MD interaction grows with particle volume as $V^2$. For Fe grains of size $6$nm
and interparticle distance $1$nm the dipole constant is $g\approx 230$ K. This value of
MD interaction equals to the peak value of exchange interaction.
The exchange interaction grows with grain surface as $V^{2/3}$ and it
is intergrain distance dependent.
\begin{figure}
\includegraphics[width=1\columnwidth]{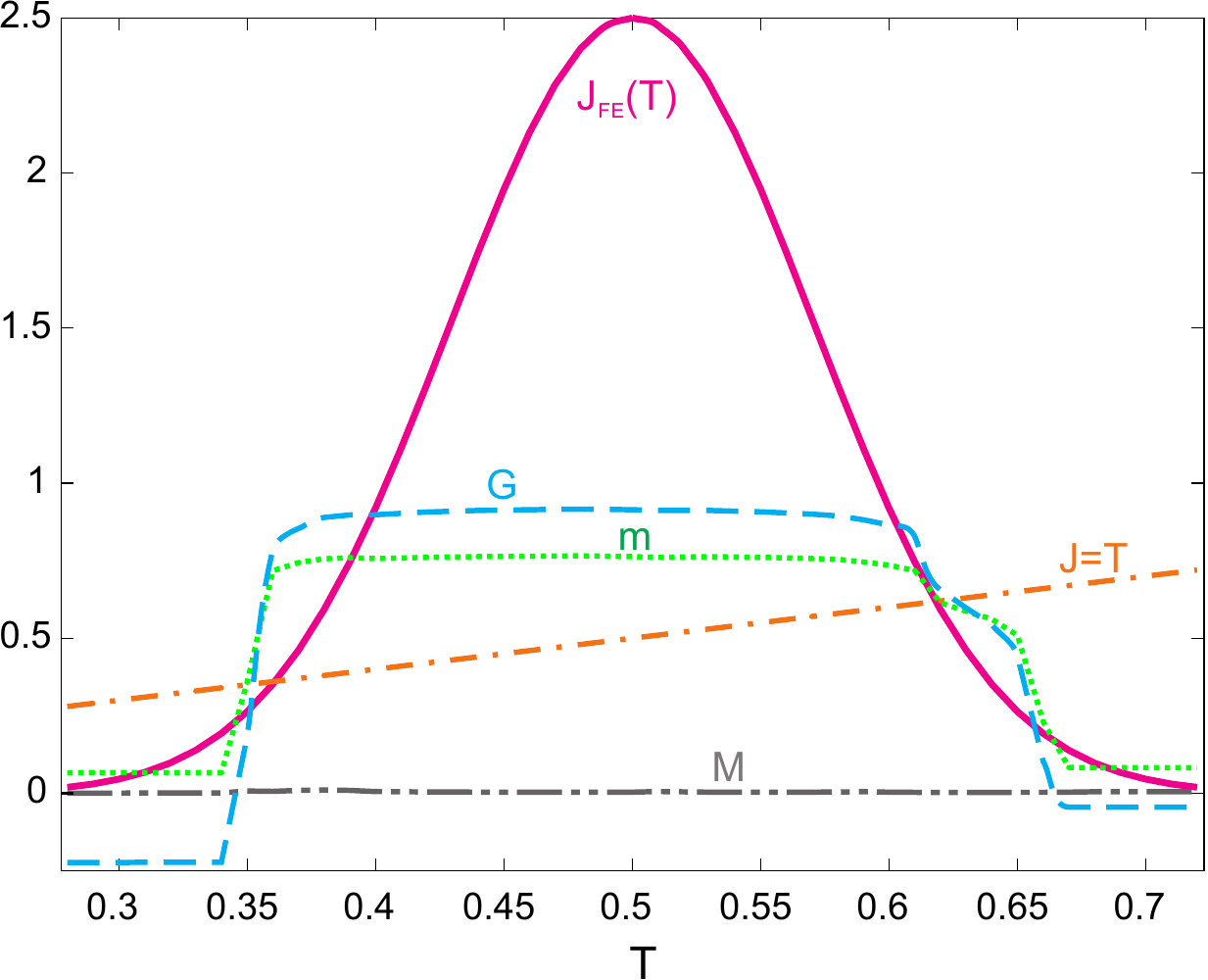}\\
\caption{(Color online) Magnetic phase diagram of composite multiferroic vs.
temperature $T$ for strong magneto-dipole interaction, $g=5.0$, and zero
magnetic anisotropy, $K=0$. All notations are defined in Fig.~1.}\label{Fig_PU_5_0}
\end{figure}

Figure~\ref{Fig_PU_5_0} shows the case of strong
MD interaction with dipole constant $g=5$ being twice larger
than the peak value of the exchange interaction.
This case is typical for $10$nm $Fe$ grains with MD
constant $g$ exceeding the room temperature.
Even in this case the FM domains exist in the vicinity of the FE Curie point. The
FM state in the vicinity of the FE phase transition is robust against the
MD interaction leading to the unusual magnetic phase diagram of composite multiferroics
due to ME coupling of ferroelectric and magnetic subsystems.

\subsection{Influence of magnetic anisotropy on the magnetic phase diagram of composite multiferroics}

In this subsection we discuss the influence of magnetic anisotropy on the magnetic phase
diagram of composite multiferroics.
The magnetic anisotropy in granular materials is much stronger than in bulk magnets
due to surface and shape anisotropy.~\cite{Chien1988,Chen1991}
It plays an important role in formation of magnetic state of granular magnets.
The magnetic relaxation time in the system of non-interacting particles
exponentially depends on the ratio of anisotropy energy and
temperature, $\tau_{\mathrm{r}}\sim \exp(K/(k_{\mathrm B}T))$.
At low temperatures the relaxation time becomes larger than the characteristic measurement time.
At these temperatures the measured magnetic properties are the properties of non-equilibrium or
``blocked'' state. The temperature hysteresis of magnetic properties is
the signature of ``blocking'' phenomenon. The Monte-Carlo (MC) calculations
in some way are similar to real experiment: simulations start with a certain non-equilibrium state and
the system ``relaxes'' to the equilibrium state via discrete steps during the simulations.
If number of MC steps $N_{\mathrm{MC}}$ (which can be associated with
measurement time if the attempt frequency of the system is known)
exceeds a certain value $N_{\mathrm r}$ (which can be associated
with relaxation time $\tau_{\mathrm{r}}$) then the system relaxes to the
equilibrium state during simulations.
In the opposite case, the system is locked into some non-equilibrium magnetic state.
\begin{figure}
\includegraphics[width=1\columnwidth]{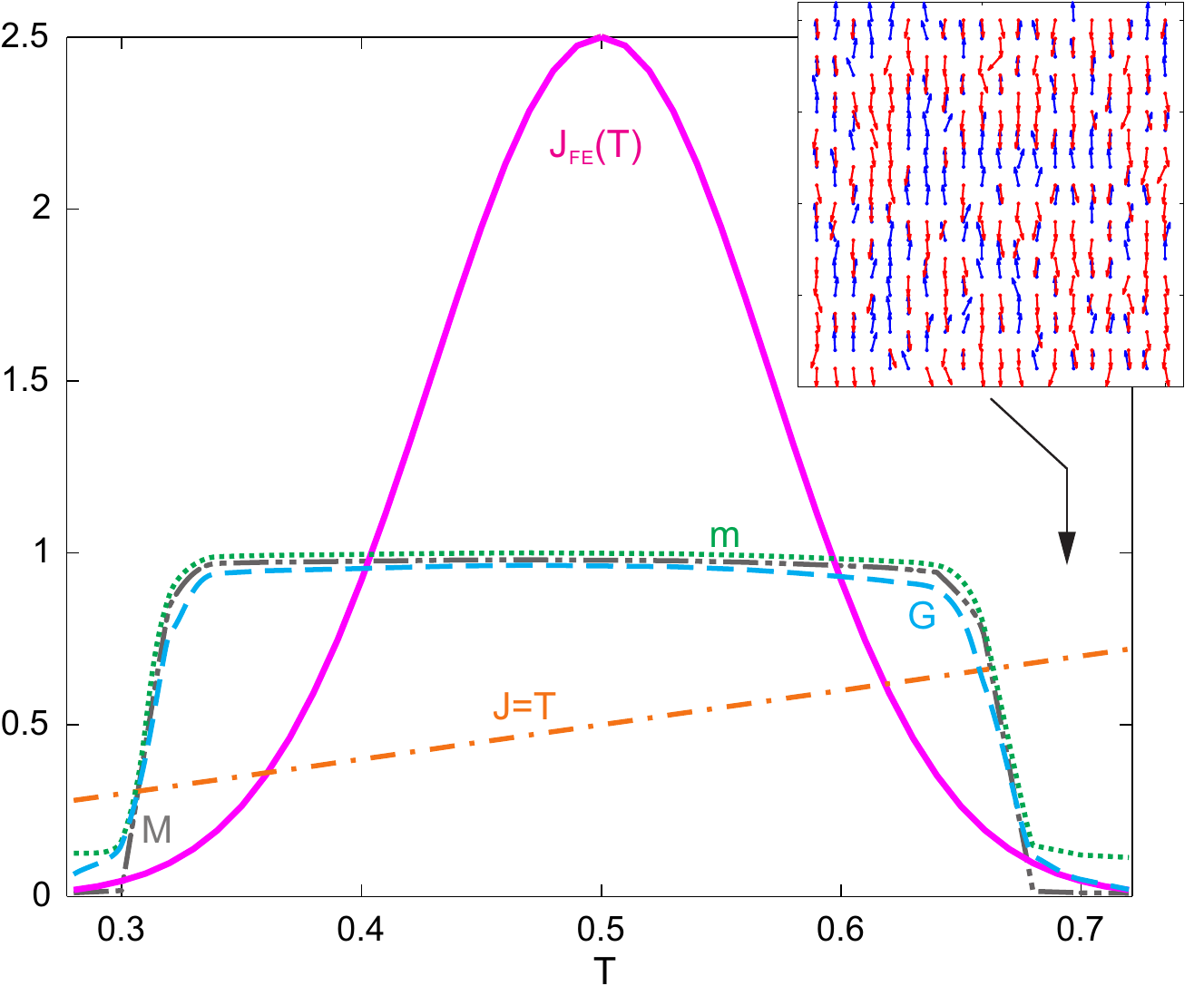}\\
\caption{(Color online) Magnetic phase diagram of composite multiferroic vs. temperature $T$
for zero magneto-dipole interaction and strong magnetic anisotropy ($K=6.0$). Insert:
Snapshot of a single magnetic layer of composite multiferroic with
disordered magnetic state due to strong anisotropy. Black arrow indicates
the position of this snapshot on phase diagram.
All notations are defined in Fig.~1.}\label{Fig_PU_6_00}
\end{figure}

Figure~\ref{Fig_PU_6_00} shows the magnetic behavior of granular multiferroics
with strong anisotropy, $K=6.0$. This anisotropy is twice larger
than the peak value of the exchange interaction. This situation is realized for $6$ nm Fe grains
with anisotropy constant $K=0.8\cdot10^{-6}$ erg/cm$^3$.~\cite{Chen1991}
Figure~\ref{Fig_PU_6_00} coincides with mean-field theory, where FM state exists
in the vicinity of the FE phase transition and the disordered state appears
away from $\TCFE$. The only difference with mean field theory is related to the fact
that magnetic moments in the disordered state have only two possible directions
along the anisotropy axis (see inset in Fig.~\ref{Fig_PU_6_00}).

For zero MD interaction the MC results do not depend on the initial
state of the system. We use two different approaches for MC calculations,
see the details in the Appendix:
1) The starting configuration is the FM alignment along the z-direction
at the initial temperature point. After a certain number of MC steps
the resulting spin configuration
is used for the next temperature point. 2) The starting configuration
is the disordered state at each temperature point. In addition,
we change the direction of temperature evolution and the number of MC steps.
Both approaches lead to the same results without hysteresis behavior. Thus,
we conclude that in our modelling $N_{\mathrm{MC}}\gg N_{\mathrm r}$.
And the magnetic anisotropy alone does not qualitatively change the
magnetic phase diagram of composite multiferroics and it does not lead to the
suppression of ME effects in the system.

\subsection{Hysteresis behavior of composite multiferroics with strong magnetic anisotropy and magneto-dipole interaction}

\begin{figure}
\includegraphics[width=1\columnwidth]{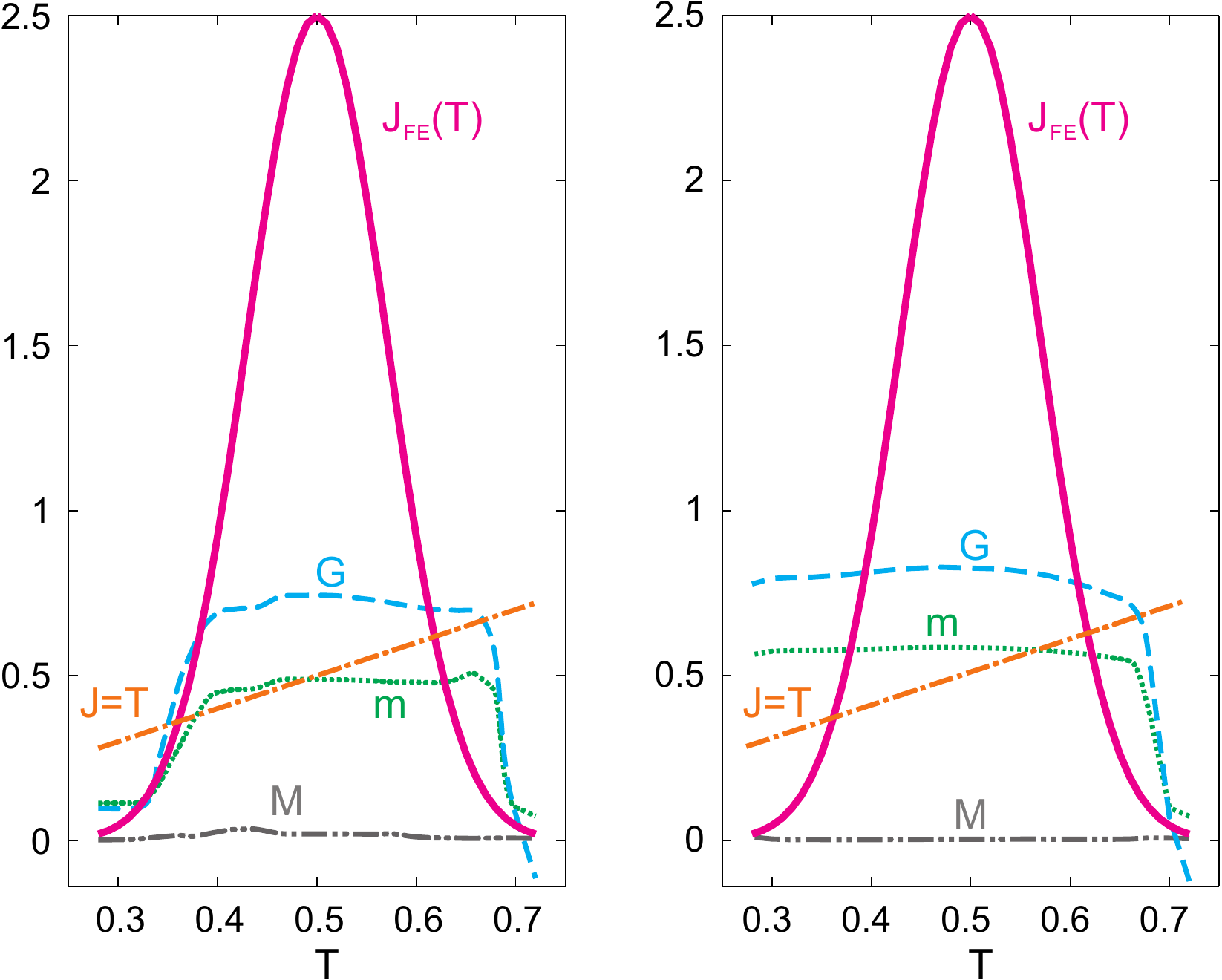}
\caption{(Color online) Temperature hysteresis of composite multiferroic. Left and right panels correspond
to the case of increasing and decreasing temperature, respectively.}\label{Fig_Hyst1}
\end{figure}

In this subsection we discuss the influence of strong magnetic anisotropy
and MD interaction on the phase diagram of composite multiferroics
and show that these interactions lead to new features.
We use the following approach: 1) We move from high to low temperatures
starting from the FM state at initial temperature $T=0.72$.
At each next temperature point we begin with the final state of the previous temperature point.
2) We move from low $T=0.28$ to high $T=0.72$ temperatures with the initial state corresponding
to the FM state. Results are shown in Fig.~\ref{Fig_Hyst1}.

The presence of MD interaction increases the blocking temperature leading
to larger magnetic relaxation time, $\tau_{\mathrm{r}}$.~\cite{Bertram2001,Maylin2000}
However, this is not the case for our system.
To understand the nature of hysteresis in composite multiferroics consider the
starting point $T=0.28$ in the ``warming'' case, left panel, with the uniform FM state as the initial state.
The final state at this temperature obtained after MC simulations is the disordered (SPM) state.
The system relaxes from FM to SPM state in the process of calculations.
Therefore neither magnetic anisotropy nor the MD interaction produce ``blocking''
in the system at these temperatures.

However, if we move from high to low temperatures the relaxation to the SPM state
does not occur at temperature $T=0.28$. The reason for
the cooling process is related to the fact that
one needs to pass the peak in the exchange interaction at temperature $T=0.5$ where
multiple inhomogeneous FM state with domains of different orientations is formed.
Around $T=0.5$ the average magnetization (the red curve) is zero.
But the cell averaged magnetization is finite meaning that
the system is divided into domains of opposite magnetization.
These domains occur due to the
interplay of exchange and MD interactions. This multidomain state is
robust against thermal fluctuations for $T<0.5$ even for zero exchange interaction.
This is in contrast to the uniform FM state,
which can be destroyed in the absence of exchange interaction even at low temperatures.
As a result the magnetization vs. temperature has a hysteresis behavior.

At higher temperature, $T\approx 0.7$, the multidomain state is unstable
against thermal fluctuations at these temperatures and the hysteresis behavior is absent.
\begin{figure}
\includegraphics[width=\columnwidth]{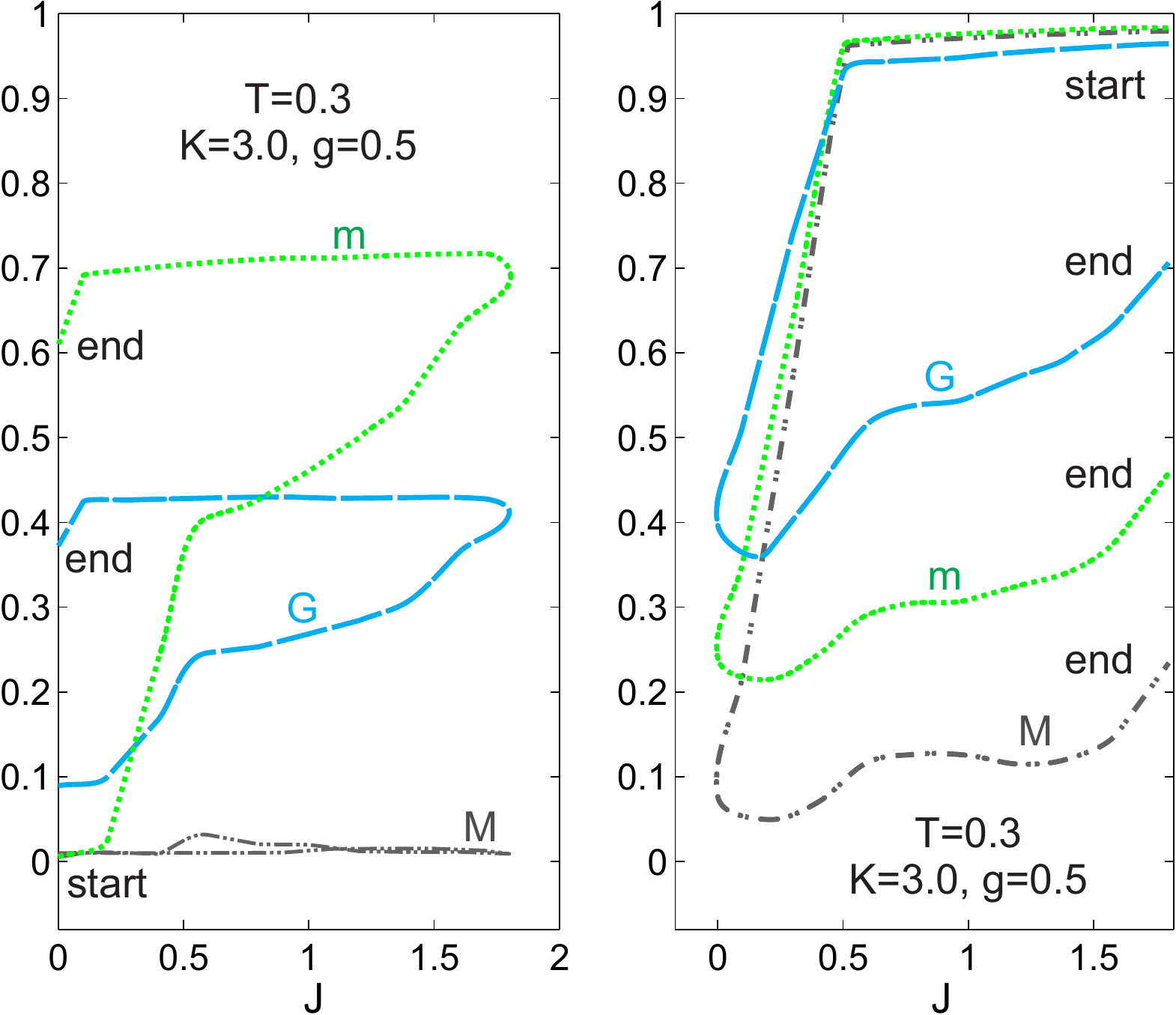}
\caption{(Color online) Evolution of magnetic phase diagram of composite multiferroic as a function
of exchange interaction at fixed temperature.}\label{Fig_Hys2}
\end{figure}

Figure~\ref{Fig_Hys2} shows the magnetic phase diagram
as a function of exchange constant $J$ at
fixed temperature $T=0.3$. On the left panel we increase the exchange constant $J$ starting from zero to
a certain high value and decrease it back to zero. The initial spin configuration at $J=0$ is
the uniform FM state. On the right panel in Fig.~\ref{Fig_Hys2} we start with finite value of
exchange interaction $J$, decrease it to zero and return back to the same value.
Obviously, such a nonmonotonic behavior of exchange interaction occurs
with changing the temperature in composite multiferroic. Figure~\ref{Fig_Hys2} shows
the hysteresis behavior caused by the non monotonic change of the
intergrain exchange interaction. Similar hysteresis occurs as a function of temperature.

To summarize, the above hysteresis is specific to composite multiferroics - materials with non-monotonic
behavior of exchange interaction. The hysteresis
is absent for systems with temperature independent exchange interaction. The
peculiar feature of this hysteresis is related to the fact that it appears in the vicinity of the FE Curie point.

\subsection{Deep in the exchange interaction in the vicinity of FE Curie point}

\begin{figure*}
\includegraphics[width=2\columnwidth]{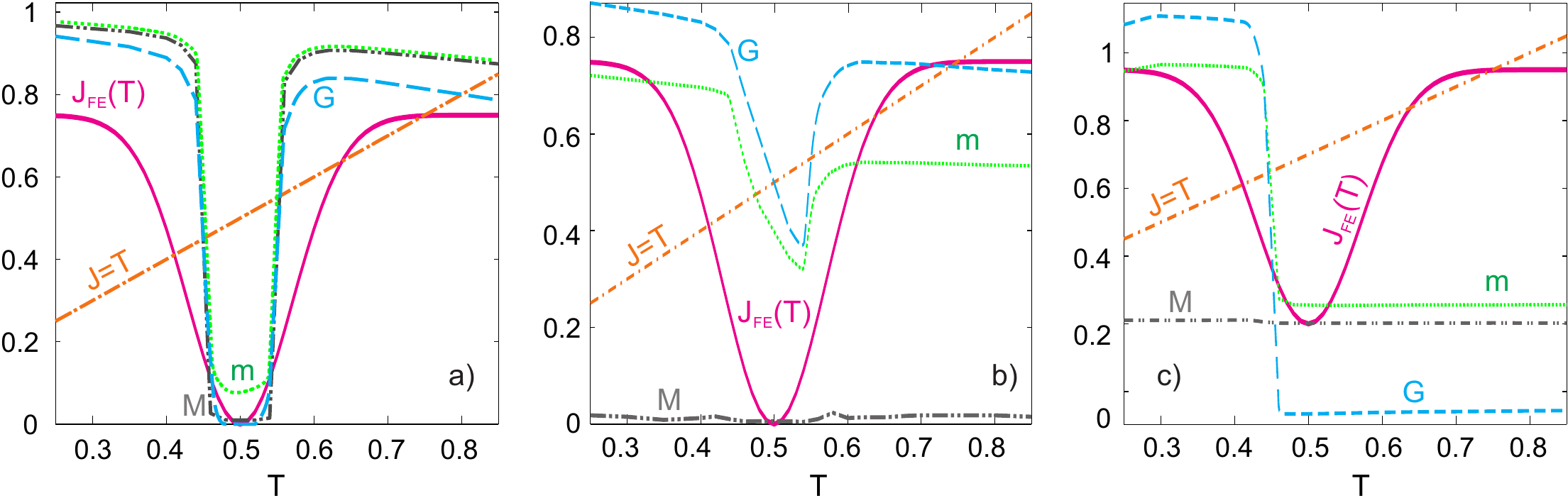}\\
\caption{(Color online) Magnetic phase diagram of composite multiferroic vs. temperature $T$
for the case of deep in the exchange interaction in the vicinity of the FE Curie point.
Panel (a): zero magneto-dipole (MD) interaction ($g=0.0$) and zero magnetic anisotropy ($K=0.0$).
Panel (b): moderate MD interaction ($g = 1.0$) and zero magnetic anisotropy ($K=0.0$).
Panel (c): strong MD interaction ($g = 1.0$) and zero magnetic anisotropy ($K=0.0$).
Solid (red) line shows the temperature dependence of intergrain exchange interaction, $J(T)$.
Straight dash dotted (orange) line stands for
temperature $T$. Gray dash dotted and green dotted lines show the average magnetization
$M(T)$ and cell averaged magnetization $m(T)$, respectively.
Blue dashed line shows the nearest neighbour correlation
function $G$.}\label{Fig_Pd_0_0}
\end{figure*}
The intergrain exchange interaction has either peak or deep in the vicinity of FE Curie point
depending on the system parameters.~\cite{Bel2014ME}
Here we study the situation with deep in the exchange interaction
in the vicinity of the FE Curie point. Figure~\ref{Fig_Pd_0_0}a shows
the case of zero MD interaction and zero anisotropy
with the following parameters: the deep is
$J_0 = 0.75$, $\DTFE=0.07$, and $\TCFE=0.5$.
In this case the Monte-Carlo simulations and the mean field approximation coincide.
For chosen parameters the exchange interaction exceeds the
temperature in the whole range except the close vicinity of
FE transition, where exchange interaction
is small and the system is in the SPM state. Outside this region the system is in
the FM state. A single domain state with two magnetic phase transitions
is realized for zero MD interaction.

A moderate MD interaction leads to the domains formation in the FM phase and
to the formation of magnetic vortices in the transition regions. Figure~\ref{Fig_Pd_0_0}b shows
this behavior for dipole constant $g = 0.5$.

Strong MD interaction ($g=5.0$) leads to the suppression of the FM state
at high temperatures, see Fig.~\ref{Fig_Pd_0_0}c. Here the AFM stripe structure appears instead
of FM ordering. At low temperatures the FM state exists.
Therefore in the case of deep in the exchange interaction the MD interaction leads to the
suppression of high temperature magnetic phase transition in contrast to the case of peak
in the exchange interaction. At higher temperatures, $T > 1$, the AFM state is suppressed
due to thermal fluctuations.

\subsection{Applicability of results}

Here we discuss the applicability of our approach. First, we study the case of regular magnetic array
with fixed intergrain distances. In real materials this distance fluctuates leading
to the dispersion of MD and exchange interactions.

Second, we consider 3D multiferroic materials which
produced via bottom-up method. A different top-down fabrication, based on layer by layer growth,
is used to produce a single layer of magnetic grains. In 2D systems the influence of MD interaction on
the magnetic phase diagram is different from 3D case. This situation requires further investigation.

\section{Conclusion}

We studied the competition of magneto-dipole, anisotropy and exchange interactions
in composite three dimensional multiferroics -- materials with magnetic grains embedded into FE matrix.
The peculiarity of composite (or granular) multiferroics is related to the fact that interparticle
interaction is affected by the FE matrix. Granular multiferroics show the magneto-electric coupling effect. Using Monte Carlo simulations we showed
that magneto-dipole interaction does not suppress the
ferromagnetic state caused by the interaction of the ferroelectric
matrix and magnetic subsystem. Thus, MD interaction does not suppress the
ME effect in granular multiferroics. However, the presence of magneto-dipole interaction
influences the order-disorder transition: depending on the strength of
magneto-dipole interaction the transition from the FM to the SPM
state is accompanied either by creation of vortices or domains of opposite magnetization.

We showed that ``blocking phenomenon'' appears at finite magnetic anisotropy and finite
MD interaction. The temperature hysteresis loop occurs due to non-monotonic behavior
of exchange interaction vs. temperature. The origin of this hysteresis is
related to the presence of stable magnetic domains which are robust against
thermal fluctuations.

\section{Acknowledgements}

We thank Shauna Robbennolt and Sarah Tolbert for useful discussions. I.~B. was supported by NSF under Cooperative Agreement Award EEC-1160504, NSF Award DMR-1158666 and the U.S. Civilian Research and Development Foundation (CRDF Global). A.~B. and N.~C.
are grateful to Russian Academy of Sciences for the access to JSCC and ``Uran'' clusters and Kurchatov center for
access to HCP supercomputer cluster. A.~B. acknowledges support of the Russian foundation for
Basic Research (grant No.~13-02-00579). N.~C. acknowledges Laboratoire
de Physique Th\'{e}orique, Toulouse and CNRS for hospitality and Russian Fundamental
Research foundation (grant No.~14-12-01185) for support of supercomputer simulations.

\appendix
\section{Calculation procedure}\label{App:CalcProc}

We use classical Monte Carlo (MC) simulations and the standard Metropolis algorithm
to model magnetic properties of the system.~\cite{Landau2000, Kretschmer79, Binder76, Binder69, Landau81, Freitas2006}
We consider $L \times L \times L$ ($L= 20$) cubic lattice with periodic boundary conditions.
To efficiently evaluate the long-range MD interaction in systems with relatively
small number of particles (as, for example, $L= 5, 6, 7$ considered in
Ref.~\cite{Kretschmer79, Kechrakos98}) one has to implement Ewald summation
technique.~\cite{Allen87, Wang01} We account the MD interaction by
direct summation in the real space applying the minimum image convention.~\cite{Allen87} In terms of the range of the interaction
that have been taken into account, this scheme is equivalent to the fast Fourier
transform method used for micromagnetic simulations.~\cite{Hinzke2000, Yuan92}

We use the FM state ordered along the $z$-direction, ${\bf S}_i= {\bf e}_z$,
as an initial spin configuration for simulating at the first temperature point.
The resulting spin state is used as an initial state
for the next temperature point and so on. To study hysteresis effects we make
two passages: first, we start with low temperature and increase the
temperature during our calculations; second, we do the opposite.

One MC step consists of $L \times L \times L$ consecutive changes in the lattice spin orientations.
We calculated the change in the energy of the system $\Delta {\cal H}$:
if change is negative, $\Delta {\cal H} \le 0$, a new state is accepted;
if change is positive, $\Delta {\cal H} > 0$, the new state is accepted with
probability $e^{- \Delta {\cal H}/T}$.
In our simulations we use $N_{MC}= 12000$ MC steps per spin and
study 60 samples in every 200 MC steps to calculate thermal properties.
To check the stability of final configuration on the number of MC steps
we increased the number of MC steps five times (up to $N_{MC}= 6 \cdot 10^4$) and find no difference in
the resulting state.

We generate the update in spin directions using two ways. First, the spin
orientations were distributed uniformly over the unit sphere's surface~\cite{Wang01}
\begin{equation}\label{update2}
  \cos \theta_i= \xi, \quad \varphi_i= \pi \xi',
\end{equation}
where $\xi, \xi'$ are some random numbers from the interval $(-1, 1)$.
This algorithm becomes inefficient at low temperatures or strong anisotropy.
In this case the majority of randomly chosen spin directions has to be rejected
due to large energy change $\Delta {\cal H}$.
We use such an update to check the results of the second algorithm
with tuned step of change in the spin direction.

Our main algorithm for spin change was an algorithm where
a new spin direction is chosen within a small angle near a given spin ${\bf S}_i$~\cite{Serena93}.
First, a random unit vector ${\bf w}$ perpendicular to the chosen spin ${\bf S}_i$ is generated.
Then new trial configuration is chosen as
\begin{equation} \label{cone}
{\bf S}'_i= \cos \theta_i {\,} {\bf S}_i+ \sin \theta_i {\,} {\bf w},
\end{equation}
where $\theta_i$, the rotation angle from ${\bf S}_i$ to ${\bf S}'_i$, is chosen according to
\begin{equation}
\cos \theta_i= 1+ \xi (\cos \theta_{max}- 1),
\end{equation}
$\xi$ is a random number varying in the interval $(0, 1)$, the angle $\theta_{max}$ is a maximum allowed amplitude for the change of the polar angle $\theta_i$ of the initial spin ${\bf S}_i$, $0 < \theta_{max} \leqslant \pi$.
The new spin direction ${\bf S}'_i$ lies within a cone around the initial direction with aperture angle $2\theta_{max}$ and all the directions inside this cone can be reached with the same probability \cite{Serena93}. The value of $\theta_{max}$ is adjusted after one full MC sweep over the lattice to keep, when possible, the number of accepted spin changes around $\sim 50\%$. Also we kept the lower bound for $\theta_{max} \gtrsim \pi/6$ to prevent too small MC moves which are inefficient to thermalize the system.

This algorithm is not valid at low temperatures~\cite{Freitas2006} or
strong anisotropy, $K \gg 1$, when the system tends to the Ising limit which
does not allow the spin flips.
We improve the situation allowing spins to flip with a certain small probability ($\lesssim 0.1 - 0.2$).
With this modification we reproduce the correct values of the critical temperature $T_c$ for
the Heisenberg, $T_c \simeq 1.44$,~\cite{Lau89, Holm93, Peczak91, Chen93, Watson69, Landau81, Binder69, Binder76, Watson69} and the Ising, $T_c \simeq 4.51$, models.~\cite{Binder01}

\bibliography{NumericalMod}

\end{document}